\documentclass[amsmath,superscriptaddress,amsmath,amssymb, aps,longbibliography,
twocolumn, a4paper]{quantumarticle}
\pdfoutput=1
\usepackage{graphicx}
\usepackage{braket}
\usepackage{dcolumn}
\usepackage{bm}
\usepackage{hyperref}
\usepackage[T1]{fontenc}
\usepackage{xcolor}
\usepackage{subfigure}
\usepackage{appendix}
\usepackage[english]{babel}
\usepackage{natbib}
\usepackage[normalem]{ulem}
\begin{document}
\title{%Thermalization in Trapped Bosonic Systems with Disorder
%Participation Ratio as a Diagnostic for Thermalization in Disordered Bosonic Chains\\
Diagnosing Thermalization via Participation Ratio in Disordered Bosonic Chains
}

\author{Javier de la Cruz}
\affiliation{Instituto de Ciencias Nucleares, Universidad Nacional Aut\'onoma de M\'exico, Apdo. Postal 70-543, C.P. 04510  CDMX, Mexico}

\author{Carlos Diaz-Mejia}
\affiliation{Instituto de Ciencias Nucleares, Universidad Nacional Aut\'onoma de M\'exico, Apdo. Postal 70-543, C.P. 04510  CDMX, Mexico}
\author{Sergio Lerma-Hern\'andez}
\affiliation{Facultad de F\'isica, Universidad Veracruzana, Campus Arco Sur, 
Paseo  112, C.P. 91097 Xalapa, Veracruz, Mexico}
\author{Jorge G. Hirsch} 
\affiliation{Instituto de Ciencias Nucleares, Universidad Nacional Aut\'onoma de M\'exico, Apdo. Postal 70-543, C.P. 04510  CDMX, Mexico}

%\date{\today}

\begin{abstract}
We study thermalization in a disordered one-dimensional interacting bosonic system described by the Aubry-André model using full exact diagonalization. We find a broad chaotic energy window where the system's eigenstates satisfy the Eigenstate Thermalization Hypothesis (ETH), demonstrated by the smooth energy dependence of observables like entanglement entropy and local particle number, whose fluctuations decrease with system size. Dynamically, we investigate the equilibration of initial Fock states and find that thermalization is not universal. The key finding is a direct and nontrivial correlation between an initial state's delocalization in the energy eigenbasis—quantified by the Participation Ratio (PR)—and its subsequent equilibration. States with a high PR consistently evolve toward the microcanonical ensemble prediction, whereas those exhibiting  a low PR display deviations whose magnitude inversely correlates with the PR value. 
This connection is quantitatively confirmed by the trace distance, providing a powerful, experimentally relevant diagnostic for predicting which initial states will reach thermal equilibrium.
\end{abstract}

\maketitle

\large

\section{\label{sec:level1}Introduction}

In recent years, experiments with ultracold atomic systems confined to one-dimensional (1D) geometries have emerged as a frontier in quantum statistical mechanics \cite{Stoferle(2004),Fallani(2007),DErrico(2014),Kaufman(2016),Lukin(2019)}. The properties of these systems, governed by the laws of quantum mechanics, present profound challenges to our traditional understanding of statistical ensembles and equilibrium states. These experiments, often conducted at ultra-low temperatures near absolute zero, allow researchers unprecedented control over individual quantum states and offer a powerful testbed for probing thermalization in isolated quantum systems.\cite{Greiner(2002),Bloch(2008),Endres(2016)}. Despite the remarkable progress in experimental techniques and theoretical frameworks, fundamental questions persist regarding the applicability of conventional statistical ensembles in describing the behavior of these few-particle systems\cite{Roux(2008),Rigol_2015}.

Thermalization processes in quantum systems have witnessed significant breakthroughs, leading to the formulation of the Eigenstate Thermalization Hypothesis (ETH), a compelling framework to explain thermalization in closed quantum systems \cite{Deutsch(1991),Srednicki_94,Rigol_2008}. It posits that eigenstates of typical quantum Hamiltonians exhibit statistical properties similar to  those predicted by the microcanonical ensemble, fundamentally implying that the system's behavior can be effectively described by equilibrium statistical mechanics. This work examines the validity of ETH in the interacting Aubry-André model (IAA), a paradigmatic disordered bosonic system that exhibits both localized and chaotic regimes depending on energy and disorder strength.

A comprehensive understanding of thermalization in closed quantum systems necessitates careful consideration of  the initial states. The approach toward thermal equilibrium is deeply influenced by the structure  
of the state from which the system evolves. In this context, Fock states—characterized by well-defined particle number distributions and valued for   their relative ease of preparation in ultracold atom experiments—provide a practical and insightful framework for probing the role of initial states in the equilibration process\cite{Pausch_2025_seed_erg}. 
 Motivated by this perspective, our study investigates the long-time dynamics of the whole set of Fock states. 
By analyzing the evolution of such states across a disordered bosonic chain, we aim to bridge theoretical predictions with experimental accessible observables. 
Concretely, we employ occupation probabilities and other local observables to assess the onset or breakdown of thermalization. We demonstrate that the degree of localization of the initial state in the Hamiltonian eigenbasis serves as a concrete criterion for predicting whether a given state will thermalize. This criterion offers valuable guidance for experimental investigations into quantum equilibration.

The paper is structured as follows: In Section~\ref{sec:model}, we introduce the Hamiltonian of the system, discuss its key features, and analyze quantum signatures of chaos in its energy spectrum. We also describe the occupation (Fock) basis, which serves as the set of initial states for our study. In Section~\ref{sec:eth_validity}, we examine three observables commonly measured in experiments—the occupation probabilities, the entanglement entropy, and the local particle number at the center of the chain—and assess their consistency with the Eigenstate Thermalization Hypothesis (ETH). Section~\ref{sec:dynamical fock} presents a dynamical analysis of the long-time behavior of Fock states, comparing the equilibrium values of these observables with predictions from the microcanonical ensemble. 
 We also  highlight the correlation between the participation ratio and thermalization. In Section~\ref{sec:lack therm}, we confirm the breakdown of thermalization for certain states using the trace distance as a diagnostic. Our conclusions are provided  in Section~\ref{sec:conclusion}.

\section{\label{sec:model}The Interacting Aubry-Andr\'e model (IAA)}

The  Interacting Aubry-Andr\'e model (IAA)  describes the dynamics of $N$ spin-less bosons on a lattice with $M$ sites in a one dimensional chain with open boundary conditions. The Hamiltonian of this system, with $\hbar= 1$, is \cite{Lukin(2019)}

\begin{align}\label{eq:Hamiltonian}
\hat{H} &= -J \sum_{\braket{i,j}=1}^M (\hat{b}_{i}^{\dagger}\hat{b}_{j}+h.c.)  + \frac{U}{2} \sum_{i=1}^M\hat{n}_{i}(\hat{n}_{i}-1) \nonumber \\
& \qquad + W \sum_{i=1}^M \cos(2\pi \beta i + \phi) \hat{n}_i.
\end{align}

The operators $\hat{b}^{\dagger}_i$ and $\hat{b}_i$ are the creation and annihilation operators of one boson on the site $i$, respectively. The first term describes the coherent tunneling coupling between neighboring lattice sites with rate $J$. The second term represents the interaction between bosons, depending on the occupancy, $\hat{n}_i=\hat{b}_i^\dagger \hat{b}_i $, in the same site $i$. The last term introduces a site disorder, with an incommensurate frequency $\beta$, phase $\phi$, and disorder parameter $W$. The Hamiltonian parameters $J$, $W$ and $\phi$ can be controlled by the experimentalists, nevertheless, the phase $\phi$ can vary randomly from one realization to another \cite{Lukin(2019)}.

The dimension of the Hilbert space of the system is determined by the expression Dim$=D(N,M)=(N+M-1)!/N!(M-1)!$, where $N$ and $M$ are the number of bosons  and sites, respectively. In this study, considering a unit filling of the lattice ($N/M=1$), we investigate four different lattice sizes: $M=7,8,9,10$, corresponding to Hilbert space dimensions of $1716$, $6435$, $24310$, and $92378$, respectively. The dynamics of the observables are calculated using full exact diagonalization. Each system size is sampled in 40 realizations by considering random values of $\phi$ in  the range $[0, 2\pi)$, following the approach outlined in \cite{Zhang}. Most observables are calculated as averages over these $40$ realizations with random $\phi$, in order to highlight some quantum phenomena

When the disorder parameter $W$ in the Hamiltonian (\ref{eq:Hamiltonian}) is significantly larger than $J$ and $U$, it disrupts the thermalization process and many-body localization (MBL) occurs. This results in the cessation of transport and bosons become localized. The transition to this localized state is highly dependent on the parameter $\beta$, which is chosen ideally as an irrational number. In practical experiments, $\beta$ is frequently associated  to the wavelengths of a pair of lasers. In this study, we employ a decimal approximation to the golden ratio $\beta=1.618$, whereas for the remaining parameters we consider  $U=4/(N-1)$, $J=1/2$, and $W=0.6$\cite{Lukin(2019)}. These parameter values are associated with the presence of chaos and a richer dynamics in the system, as previously observed in studies such as those of Refs. \cite{Kolovsky_2004,Kollath_2010,delaCruz(2020),DiazMejia(2024)}. 

Notice that the largest value that the interaction term can have occurs when all bosons are localized at the same site, being proportional to $N(N-1)$. The selected scaling for the interaction parameter $U= 4/(N-1)$ guarantees that this maximum value scales with $N$, as do the other two terms in the Hamiltonian, allowing a meaningful comparison of observables calculated for different system sizes\cite{pausch2022optimal,Pausch_2025_seed_erg}.

\subsection{\label{sec:rmt}
Spectral signatures of quantum chaos}

Commonly used indicators of quantum chaos rely on the properties of the energy spectrum, and determine how  close is this spectrum  to that of an ensemble of  random matrices. For a sorted set of eigenenergies, denoted as  $E_n$ (with $n = 1, 2,...,\text{Dim}$), the spacing between nearest neighbors, $s_n = E_{n+1}-E_n$, show two different distributions.
Quantum chaotic systems exhibit level spacing that follows the Wigner-Dyson distribution. 
 For  the Gaussian Orthogonal Ensamble (GOE), it is \cite{Stockman_2000}
$
    P(\tilde s)= \frac{\pi \tilde s}{2%\sigma^2
    } e^{-\tilde \pi s^2/4%\sigma^2
    }$,
in terms of the unfolded variable $\tilde s = s/\Delta E$, being $1/\Delta E$ the  density of states.
Conversely, in regular systems where eigenvalues are not correlated, the distribution of level spacing aligns well with the Poisson distribution $p(\tilde s) = e^{-\tilde s} $.

In order to avoid the unfolding procedure, one can use the ratio of two consecutive level spacings, a method proposed by Oganesyan and Huse\cite{Oganesyan_2007,Chavda_2014,Herrera_2020},
\begin{equation}\label{ratio1}
\tilde{r}=\frac{\text{min}(s_n,s_{n-1})}{\text{max}(s,s_{n-1})}=\text{min}\left( r_n,\frac{1}{r_n} \right)
\end{equation}
with $r_n=s_n/s_{n-1} $. This ratio is independent of the  density of states and, as said before, no unfolding is necessary. For quantum chaotic systems, the mean level spacing ratio is $\langle \tilde{r} \rangle_W = 4-2\sqrt{3} \approx 0.535$, whereas for regular systems, it has a lower mean value of $\langle \tilde{r} \rangle_P = 2 \ln 2 - 1 \approx 0.386$ \cite{Atas_2013,Srivastava_2018}. 
\begin{figure*}
    \centering
    \includegraphics[scale=0.9]
{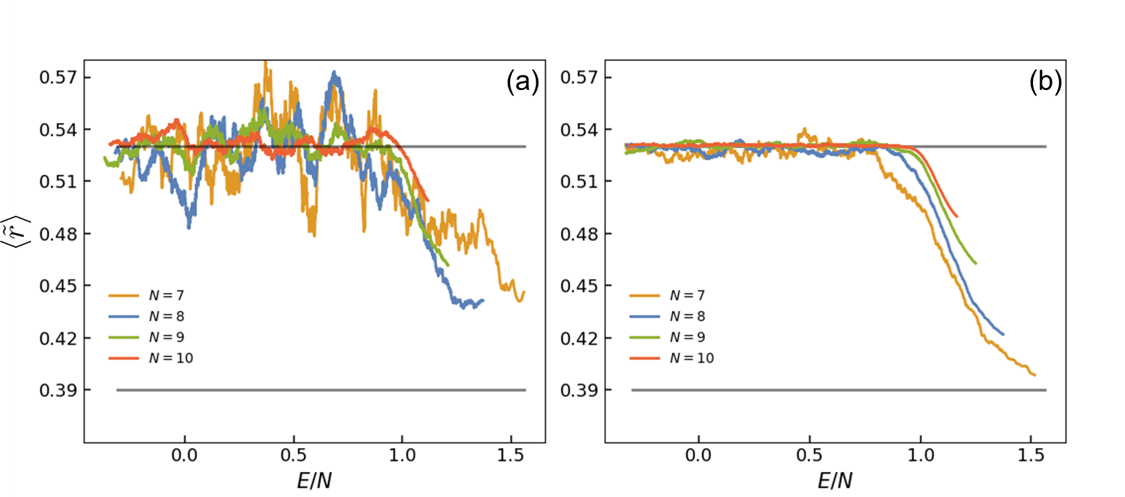} 
    \caption{Averaged level spacing ratio $\braket{\tilde{r}}$ for various  system sizes.  Horizontal lines indicate the reference values  $r_W=0.53$,  and $r_P=0.38$,   corresponding   Wigner-Dyson and Poisson statistics,  respectively. Panel (a) shows results for a single  random realization of the  phase $\phi$, using a moving window containing $0.07$ Dim states. Panel  (b)  display   averages over 40 disorder realizations,  computed using a moving window of 200 states. The Hamiltonian parameters are $\beta=1.618$, $U=4/(N-1)$, $J=1/2$, and $W=0.6$.}
    \label{fig:r_promedio}
\end{figure*}

The effect of the disorder term with intensity $W$ has been studied in detail in our previous work, Ref. \cite{DiazMejia(2024)}. It was shown that, when $U\approx J\approx 1/2$, chaos is present in the region $W\leq 1$ for different values of $N$ and $M$ for all the cases considered. In Fig. \ref{fig:r_promedio}(a) the mean level spacing ratio $\langle \tilde{r}\rangle$ is plotted as a function of the energy for  the Interacting Aubry-Andre model (IAA) in only one random realization, for $N=M=7, 8, 9$ and $10$ and Hamiltonian parameters $U=4/(N-1)$, $J=1/2$,  $W=0.6$ and $\beta=1.618$.  For the low energy sector, the parameter $\langle \tilde{r}\rangle$ oscillates around the value 0.53, indicating the presence of spectral chaos,  and, similarly to the Bose-Hubbard (BH) model,  the fluctuations diminish  as the dimension of the system increases. However,  in contrast  to the BH model,  the disorder present in the IAA model induces  localization in the highest energy eigenstates. In this regime,  bosons become confined to a single well, and both inter-well tunneling and chaos   are  significantly  suppressed. This phenomenon at high energies, % resembles to Anderson localization \cite{Teichert_2014} and  
known as self-trapping \cite{Smerzi_1997, Milburn_1997,Longhi_2011,Fangzhao_2021} emerges  for states  that lie above  a critical energy threshold,  often referred to as the mobility edge.

Fig. \ref{fig:r_promedio}(b) illustrates the level spacing ratio $\langle \tilde{r} \rangle$ versus the mean energy of the levels in intervals of 200 energy states for the IAA model, averaged over 40 realizations and for four different system sizes. The chaotic region extends from $E/N = -0.5$ to $E/N \approx 0.8$ for $M=7$, whereas for larger system sizes the upper chaotic limit increases up to $E/N \approx 1.0$ for $M=10$.  In contrast with the previous figure, the average over many realizations erases most of the fluctuations. In all cases, for high-energy states, the system transitions to a regular spectrum, confirming that the localization introduced by the disorder is always present, independent of the random phase.

\begin{figure*}
    \centering
    \includegraphics[scale=0.9]{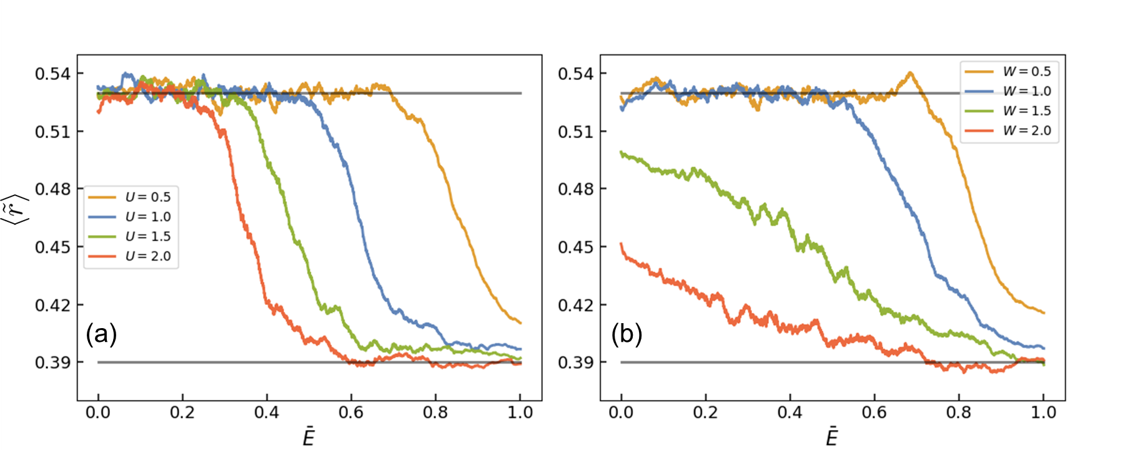} 
    \caption{Averaged level spacing ratio $\braket{\tilde{r}}$ as a function of normalized mean energy $\bar{E}=\braket{\frac{E-E_{GS}}
    {E_{\text{max}}-E_{GS}}}$, where $E_{GS}$ and $E_\text{max}$ are the ground-state and maximal energy. Panel (a)  shows $\braket{\tilde{r}}$ for different interaction strengths at fixed disorder $W=0.6$; panel (b) shows results for varying disorder strength at fixed interaction $U=0.57$. Averages are computed over rolling windows of 300 energy levels.  Horizontal lines mark reference values $r_W=0.53$,  and $r_P=0.39$  corresponding  to Wigner-Dyson and Poisson distributions, respectively. Remaining Hamiltonian parameters are $\beta=1.618$, $J=1/2$, and system size $N=M=8$.} \label{fig:r_seltrapping_1}
\end{figure*}

Finally, to reinforce our election of parameters ($U\approx J=1/2$ and $W=0.6$), in Figure \ref{fig:r_seltrapping_1}, we show the dependence  of the appearance  of chaos on the parameters $W$ and $U$. First, with $W=0.6$ held constant, panel (a) reveals that increasing the interaction parameter causes a rapid transition of   $\langle \tilde{r} \rangle$ to a regular regime at  higher energies, while   shrinking  regions of chaos persist  at lower energies. This indicates  that the mobility edge shifts toward   lower energies as the interaction strength  increases. Second, in panel (b), for a fixed interaction strength $U=0.57$, we observe that increasing the disorder parameter drives the entire spectrum into a regular regime, %$\langle \tilde{r} \rangle$ 
 a signature of  many-body localization. From this analysis, we   conclude that the parameter set  $J=1/2$, $W=0.6$ and $U=4/(N-1)$ (which corresponds to  $U\approx 0.5$ for $N=7,8,9$ and $10$) ensures a broad chaotic energy window that  spans  a significant portion of the spectrum.

\subsection{\label{sec:occupation}Characterization of Fock states}
We consider all Fock states as initial conditions, which  together form a complete basis for the systems's Hilbert space \cite{Fock1932,Reed1975}. Most of these states have mean energies concentrated near the center of the spectrum. Each state in this basis is defined by the occupation numbers $\ket{\mathbf{n}} \equiv \ket{n_1, n_2, \ldots, n_M}$, where $n_i$ denotes the number of bosons at site $i$ on a one-dimensional chain of $M$ sites. The total particle number is fixed, $\sum_k n_k = N$.

 \begin{figure}[h]
 \includegraphics[scale=0.55]{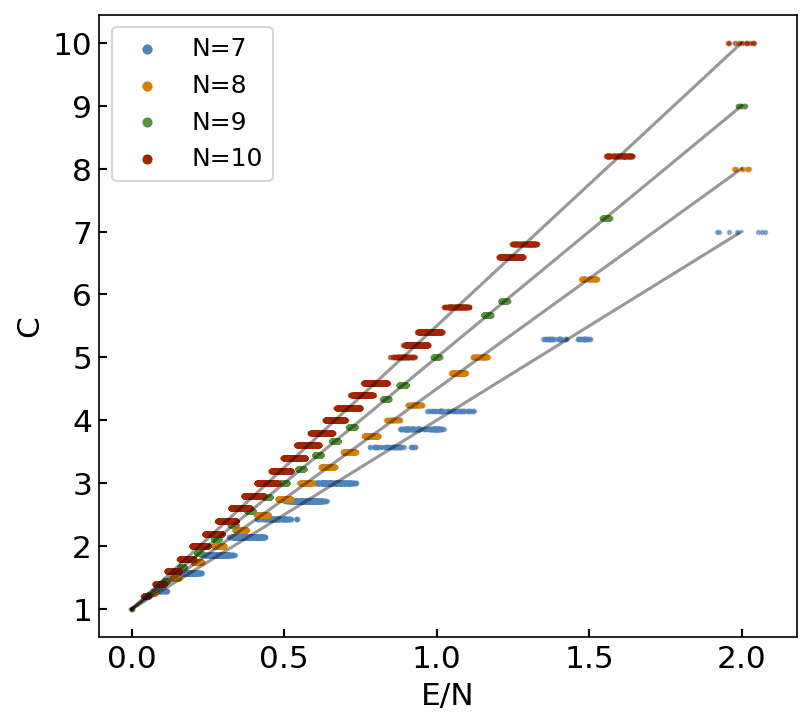}
\caption{Crowding parameter C plotted  against the mean  energy per particle $E/N$ for  Fock  states across  different system sizes and a single disorder realization. Solid  lines indicate the expected dependence of the crowding parameter for averages over disorder realizations $C=1+\frac{N-1}{2}\frac{E}{N}$.
  }\label{fig:CvsE}
 \end{figure}

To classify these states in  terms of their energy and spatial structure, we introduce the crowding parameter $C$ \cite{DiazMejia(2024)}, defined as
\begin{equation}
C = \frac{1}{N} \sum_i n_i^2.
\end{equation}
This quantity ranges from $C=1$, corresponding to the uniform Mott state ($n_i = 1$ at each site), to $C = N$, where all particles are concentrated at a single site. The crowding parameter thus  provides a measure of how spatially concentrated the bosons are in a given Fock state.

Since the  hopping term has no diagonal elements in the Fock basis, and  the disorder contributions  average out to zero  over disorder realizations, the expectation value of the energy for  Fock states is primarily determined by the on-site interaction term. This results in a linear relationship between mean energy and crowding parameter for Fock states:
\begin{align}
E = \bra{\mathbf{n}} H\ket{\mathbf{n}} &\approx \frac{U}{2} N ( C - 1) \nonumber \\
&  = \frac{2}{N-1} N ( C - 1),
\label{E(C)}
\end{align}
where in the last step we used $U=4/(N-1)$ which ensures extensivity of the maximal mean energy of Fock states, $E_{\text{max}}=\frac{2}{N-1}N(C_\text{max}-1)=2N$. Fig. \ref{fig:CvsE} shows the behavior of   $C$ as a function of the mean enegy, $E/N$, for all  Fock states and for % Each dot represents the expectation values of the energy for
a single  random realization. The mean energy average over many realization  is represented   by  straight lines given by   $C= 1 + \frac{N-1}{2} \frac{E}{N}$. 
As expected, for all four system sizes studied,  the highest energies values fluctuate around $E/N=2$. This outcome  results from the chosen  scaling for the interaction parameter $U= 4/(N-1)$.

\begin{figure*}
\centering
\begin{subfigure}
\centering
\includegraphics[scale=0.6]{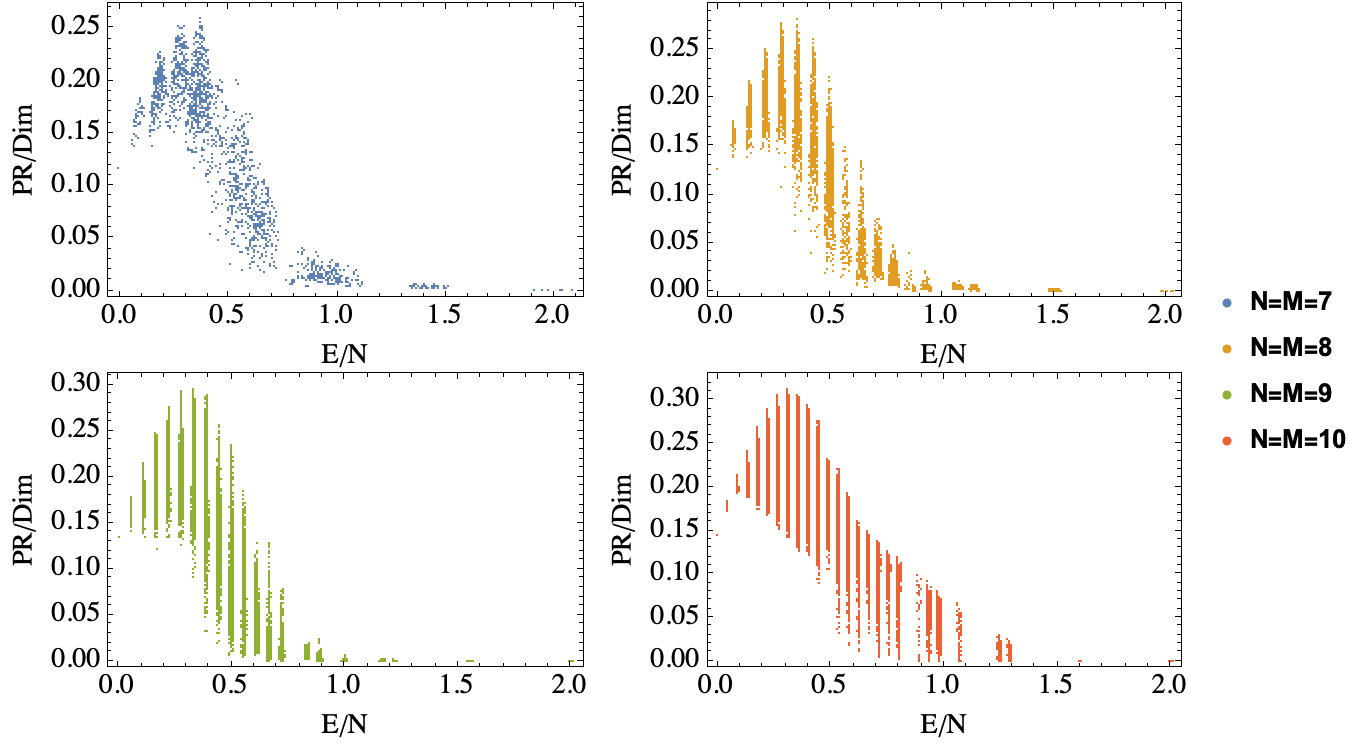}
\end{subfigure}
\caption{Normalized Participation Ratio (PR/Dim) with respect to the Hamiltonian eigenbasis, plotted against the mean energy $E/N$ normalized  for all  Fock  states of four  different system sizes.
}
\label{fig:PRE}
\end{figure*}

To characterize the structure of Fock states in the energy eigenbasis, we compute their Participation Ratio (PR),
\begin{equation}
\mathrm{PR}_\mathbf{n}= \frac {1} {\sum_m |c^{(\mathbf{n})}_m|^4} .
\label{PR}
\end{equation}
where $c^{(\mathbf{n})}_m$ are the expansion coefficients of the Fock state $\ket{\mathbf{n}}$ in the energy eigenbasis ${ \ket{\phi_m} }$. A high PR indicates that a state is delocalized across many eigenstates, while a low PR reflects localization onto a small subset.

Fig. \ref{fig:PRE} presents the PR of all occupation states in the eigenbasis, plotted versus  their average energy $E/N$. Note  that the occupation states accommodate in several clusters with approximately the same energy, these clusters are associated to the different values of the crowding parameter, which is the same clustering that  can be seen in Fig.\ref{fig:CvsE}. It can be observed that most occupation states, whose energies lie in region $E/N<0.5$ have $\mathrm{PR} > 0.10$Dim, i.e., are the more delocalized in the eigenbasis. On the other hand, those states with larger crowding parameters have the highest energies and the lowest PRs.

 Using Husimi functions in systems with two classical degrees of freedom \cite{Bastarrachea2016,Villasenor2021}, it has been shown that  localization of a quantum  state in the eigenbasis closely    correlates with its localization in phase space. Strong delocalization is typically  associated with  classical chaos, where trajectories explore the entire accessible  phase space at a given energy.  
 
 This suggest that low-energy Fock states,  which exhibit high PRs,  are  delocalized and capable of  exploring a significant portion of their energetically  accessible phase space. This behavior aligns with  expectations for  systems where interactions dominate over disorder. Conversely,  for high-energy Fock states disorder becomes more influential, leading to localization. This restricts the effective phase space  accessible to these high-energy states, resulting in a lower PR that decreases with energy more rapidly than the volume of the corresponding  energy shell. This can be seen by comparing Figure \ref{fig:PRE} with the model´s density of states shown in Figure 2(b) of Ref.\cite{DiazMejia(2024)}. 

\section{Thermalization in the IAA model}
\label{sec:therm}
The Eigenstate Termalization Hyphothesis (ETH) \cite{Rigol_2008} postulates that the expectation value $\bra{\phi_m} \hat A \ket{\phi_m} $  of a local  observable $\hat A$, evaluated in  Hamiltonian eigenstates $\ket{\phi_m}$, remains  approximately constant across   most of the eigenstates confined to  a  narrow energy interval of width $\Delta E$. This  implies that generic quantum states whose components in the eigenbasis are concentrated within this energy region will yield similar   long-time  average of that observable. 
In the following  subsection we investigate  the  validity of ETH in the model eigenstates. We focus on  three  local observables:  the occupation probabilities, 
entanglement entropy and number of bosons at the central chain site. Subsequently, we examine whether the equilibrium values of these observables - when initiated from Fock states- coincide or deviate from predictions of  the microcanonical ensemble.

\subsection{ETH validity}
\label{sec:eth_validity}

Occupation probabilities are among the most frequently measured  observables in  experiments. They represent  the probability of  finding $i$ particles at a specific site in the chain, and can be computed from the reduced density matrix.  To  mitigate   border  effects, we focus on the central site of the chain.  For a chain  of length $M$, this corresponds to  the site  indexed $k_c=\lceil M/2 \rceil$.  We consider  a bipartition of the system into  the central  site % $M_A=M/2$ (with $n_{i}$ configurations of bosons) 
and the remaining $M-1$ sites of the chain. An arbitrary pure state of the full  chain can be expressed as $$ |q\rangle=\sum_{i=0}^N\sum_{k=1}^{D_i} c_{i,\mathbf{n}_k}^{(q)} |n_1, n_2,...,n_{k_c}\!\!\!=\!\!i, ...,n_M\rangle, $$
where the array $$\mathbf{n}_k~=~(n_1,...,n_{k_c-1}, n_{k_c+1},...,n_M)$$ specifies the configuration   of bosons in the chain  excluding the central site, and $D_i$ denotes the number  of these configurations compatible with   $i$ bosons occupying  the central site.  The expression  for $D_i$ is given by 
\begin{align}
    D_i &=D(N-i,M-1)   \nonumber\\
    &=\binom{N-i+M-2}{M-2}.
\end{align} 
The reduced density matrix for the central site of the chain is given by
 \begin{equation}\label{eq:p_i_2}
    \rho_{M/2}^{(q)}=\sum_{i=0}^N \sum_{k=1}^{D_i} |c^{(q)}_{i,\mathbf{n}_k}|^2\ket{i}\bra{i} .
\end{equation}
From this reduced density matrix, %Eq.~\ref{eq:p_i_2} we can observe that 
the occupation  probability $P^{(q)}_i$ of finding $i$ bosons in the central  site is \begin{equation}\label{eq:p_i}
    P^{(q)}_i=\sum_{k=1}^{D_i} |c^{(q)}_{i,\mathbf{n}_k}|^2
\end{equation}
for  $i= 0,1,2,...,N$.

Once  the occupation probabilities are obtained, the expectation value of the particle number at the central site $\hat{n}_{M/2}$, and the  entanglement entropy $S_{M/2}$ between the central site and the rest of the chain \cite{Rigol_2015} are given respectively  by
\begin{equation}\label{eq:numberM}
n^{(q)}_{M/2}=\sum_{i=0}^N i \,P^{(q)}_i,
\end{equation}
%\,\,\,\,\,\,\,\,\, \hbox{and}\,\,\,\,\,\,\, 
and
\begin{equation}
\label{eq:shanon}
S^{(q)}_{M/2}=-\sum_{i=0}^N P^{(q)}_i \ln P^{(q)}_i.
\end{equation}

\begin{figure*}
\centering
\begin{subfigure}
\centering
\includegraphics[scale=0.68]{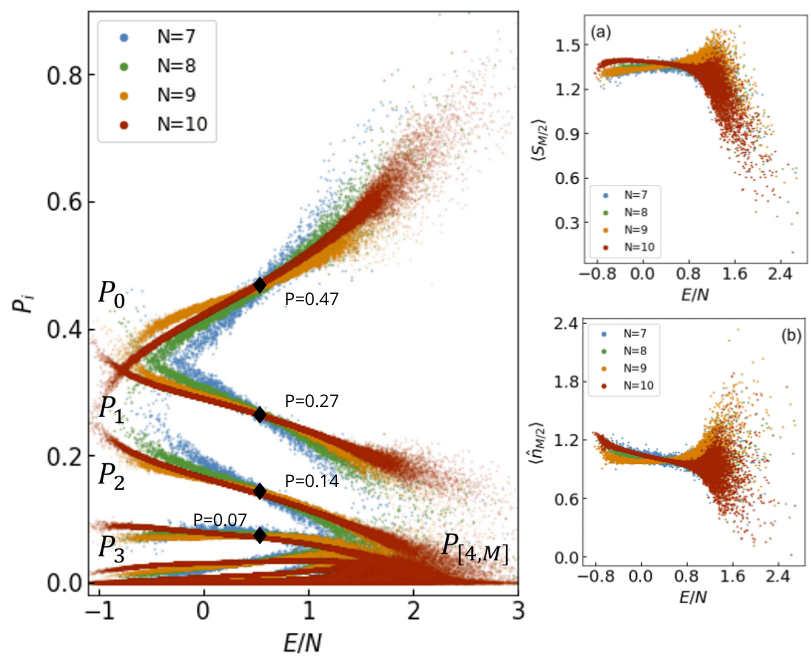}
\end{subfigure}
\caption{Left panel: Occupation probabilities at the central site of the chain for Hamiltonian eigenstates, plotted against the corresponding eigenenergies. Diamonds indicate the occupation probabilities for an infinite-temperature state, for which   $P_i^{(\text{mix})}=D_i/\sum_{i'}D_{i'}$.   Right panels:  (a) Entanglement entropy and (b) particle number at the central site for   eigenstates, both as function of corresponding eigenergies.  For all panels, results correspond to averages over 40 disorder realizations, and  four system sizes with unit filling  $N=M$. These observables exhibit smooth energy dependence and reduced fluctuation in the chaotic regime, consistent with the Eigenstate Thermalization Hypothesis.}
\label{fig:pis_eth}
\end{figure*}

The left panel of Figure \ref{fig:pis_eth} display  the  occupation probabilities at the  central  site  of the chain for all  eigenstates of the IAA model, averaged  over 40 disorder realizations.
 Across all system sizes, within the energy interval $0 \leq E/N$, the probability of finding zero bosons at the central site  %in the central site %(and also in the first site, see Supplementary Information ) 
 is dominant, and  the probability of higher occupation numbers   decreases monotonically with  particle number.
Near $E/N\approx 0.6$, the occupation probabilities resemble  a distribution characteristic of a  completely mixed (infinite-temperature) state, $\rho_\text{mix}=\mathbf{1}=\sum_\mathbf{n} |\mathbf{n}\rangle\langle \mathbf{n}|$, for which the occupation probabilities in the central site are solely  determined by  the number of configurations with $i$ bosons  $P_i^{(\text{mix})}=D_{i}/\sum_{i'=0}^N D_{i'}$, these occupation probabilities  are  indicated by diamond symbols in the Figure~\ref{fig:pis_eth}.

In the energy range $-1\leq E/N\leq 1$,  which was  shown to be chaotic in Fig. \ref{fig:r_promedio}, 
the eigenstate-to-eigenstate  dispersion of the occupation probabilities is relatively  small and  decreases  with increasing  system size. This trend  provides early evidence for  thermalization in this energy regime.
Towards the upper edge of the spectrum,  left panel of Fig.  \ref{fig:pis_eth} shows a marked  increase in  dispersion,  consistent  with the transition to a regular regime and the onset of many-body localization, as also indicated in  Fig. \ref{fig:r_promedio}.

The small dispersion of observables in the  model's eigenstates  at low-energies --and their  increased variability in the high-energy, regular   region--  is more evident   in the entanglement entropy $\braket{S_{M/2}}$ and the number of bosons  at the central site, $\braket{n_{M/2}}$, shown  in panels (a) and (b) of Fig.\ref{fig:pis_eth}, respectively. These panels  display  averages over 40 disorder realizations,  covering  all eigenstates.

For both  the entanglement entropy and  the number of bosons,  it is evident that   eigenstates with energies below approximately  $E/N \lesssim 0.8$ satisfy the Eigenstate Thermalization Hypothesis (ETH). In this energy range, the disorder averaged  observables   exhibit a smooth dependence on $E/N$, with minimal dispersion. 
In contrast,  in the higher-energy (regular)  region, the dispersion becomes significantly  more pronounced across all dimensions.

\begin{figure}
    \centering
    \includegraphics[scale=0.6]{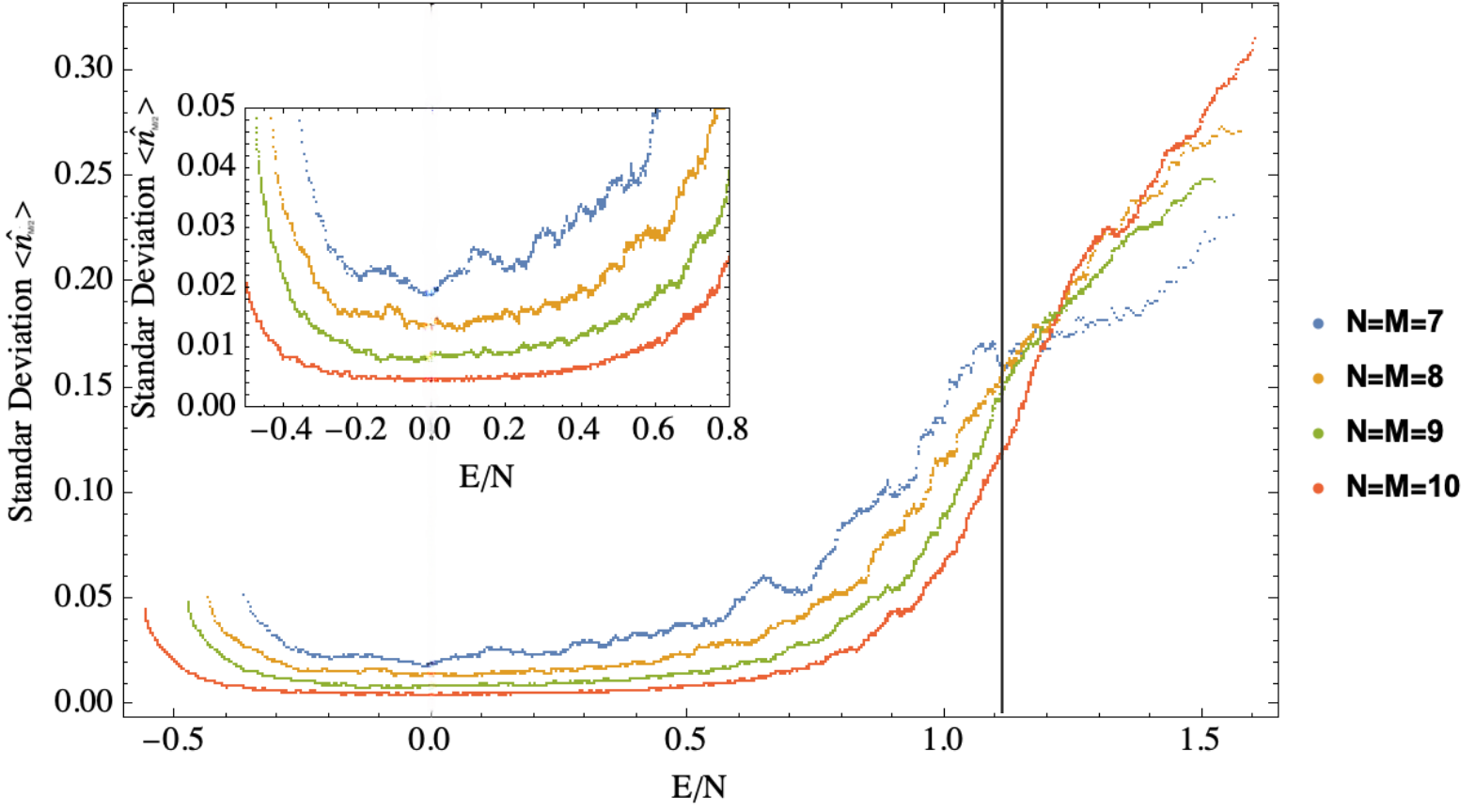}
    \includegraphics[scale=0.6]{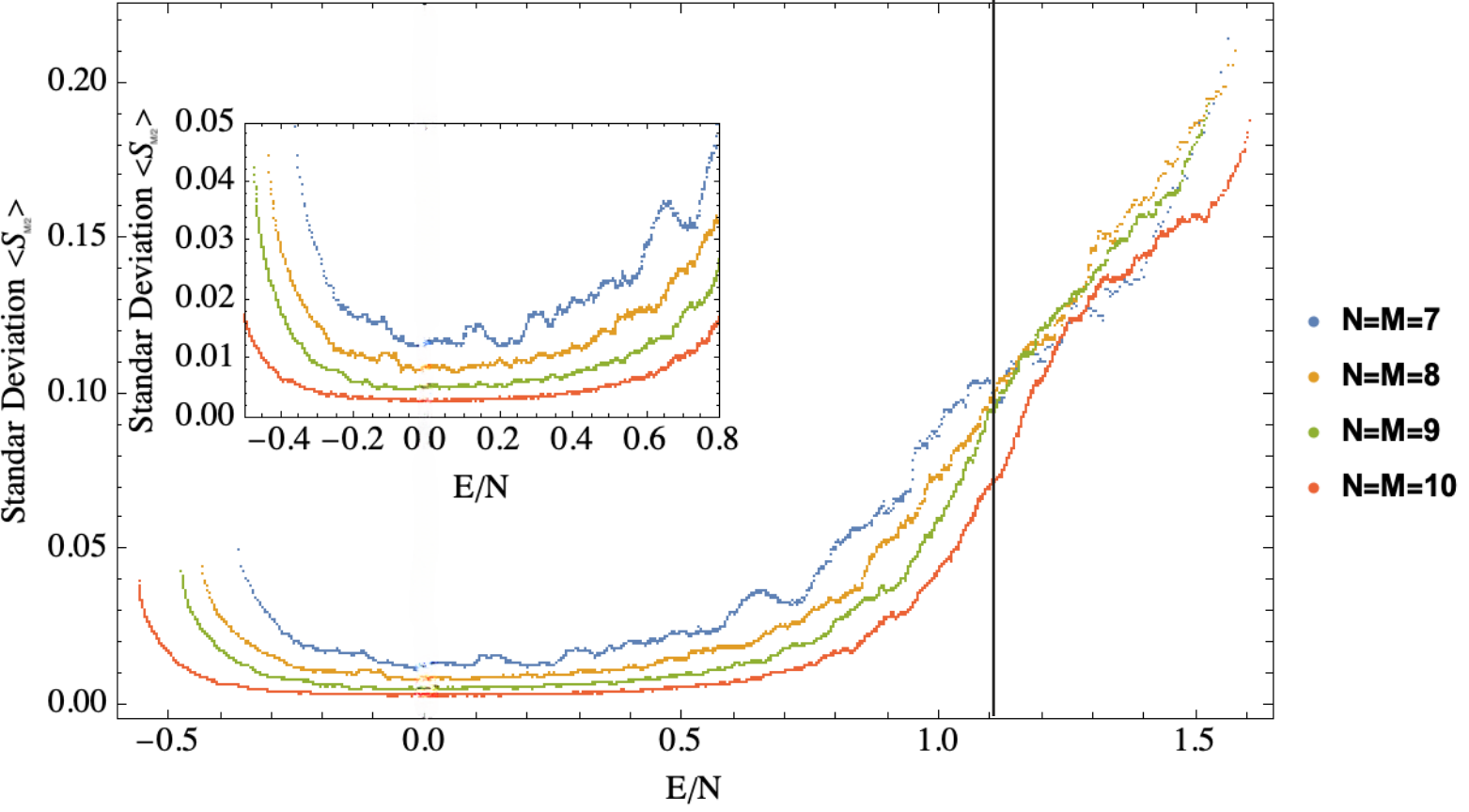}
    \caption{Top panel: Standard deviation of the disorder-averaged  particle  number at central site  of the chain, $\hat{n}_{M/2}$, computed across eigenstates using rolling windows of 200 states, and plotted as a function  of their average energy. The vertical black line near $E/N = 1.1$ marks the region where the Eigenstate Thermalization Hypothesis (ETH) ceases to hold. This indicates a transition beyond which thermalization no longer accurately describes the system's behavior. Bottom panel: same as the top panel, but  for the entanglement entropy between   the central site and the rest of the chain. Insets provide  zoomed-views of the central energy region, highlighting the decrease in  standard deviations with increasing system size.}
    \label{fig:StandarDeviation}
\end{figure}

\begin{figure}[h!]
    \centering
    \includegraphics[scale=0.6]{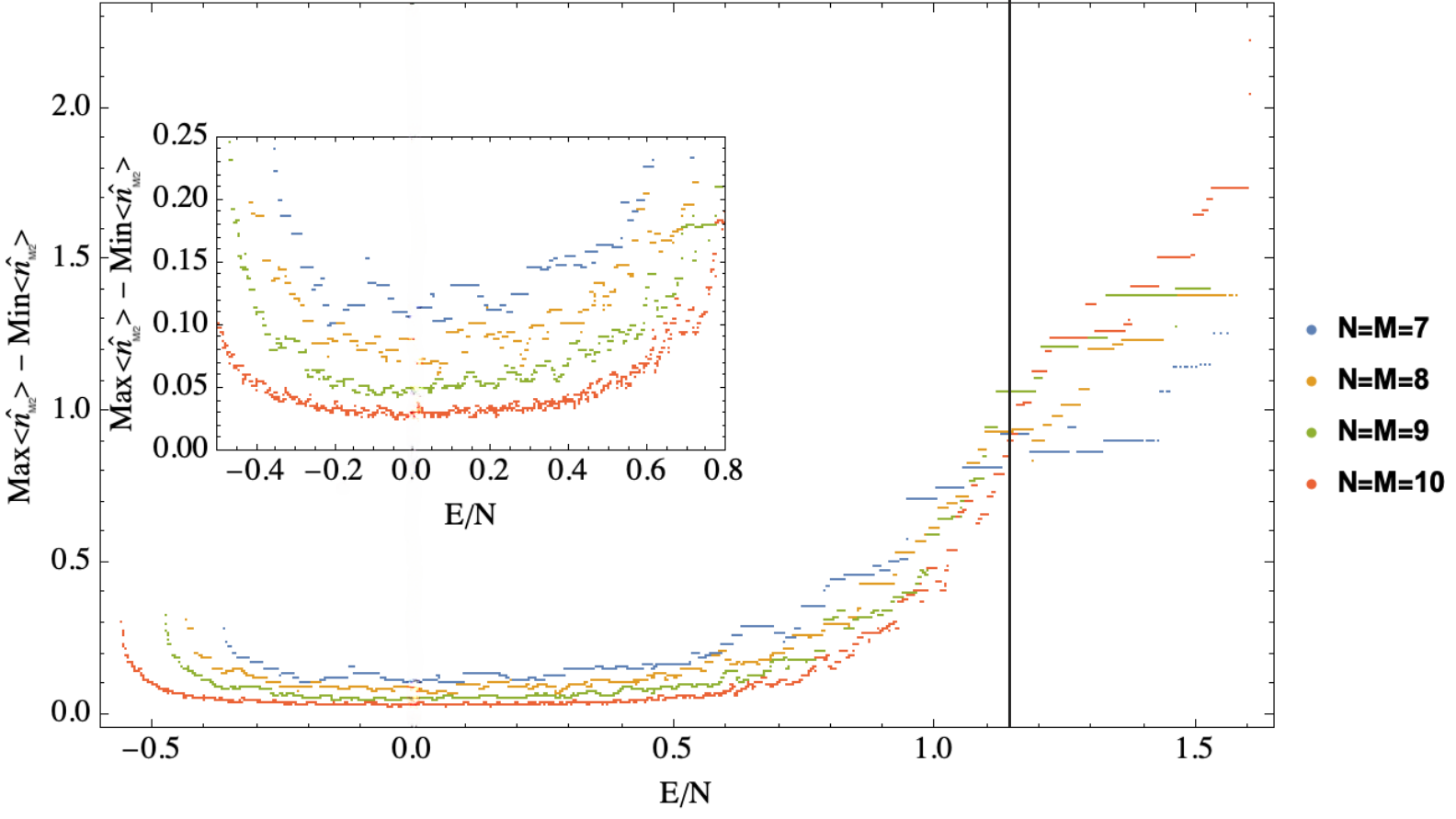}
    \includegraphics[scale=0.6]{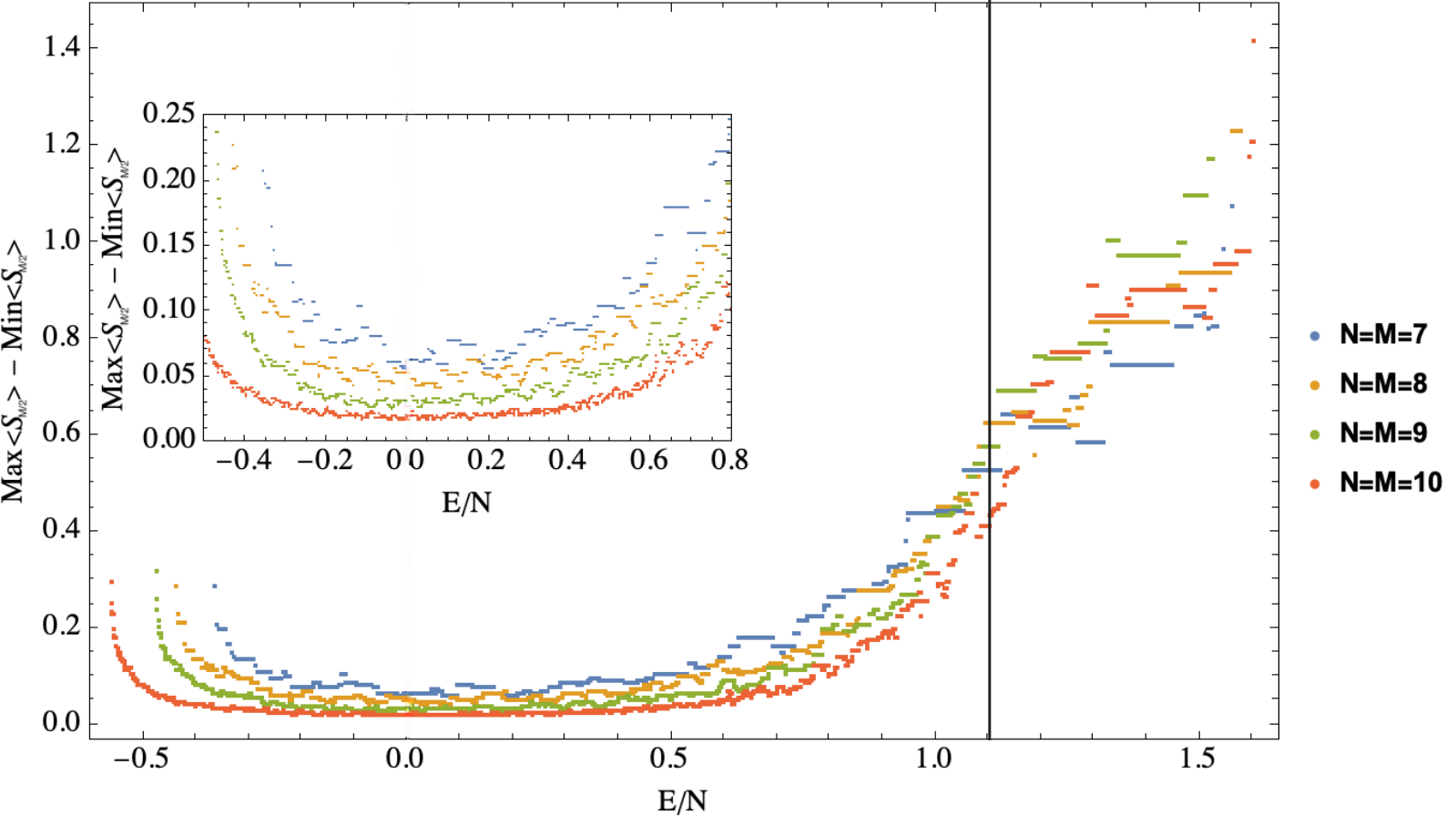}
    \caption{Top panel: Range  (maximum   minus  minimun value)  of the disorder-averaged    particle number at  the central site of the chain, computed  over  rolling windows of 200 eigenstates and  plotted against  the mean energy of each window.  The vertical black line near $E/N = 1.1$ marks the region where the Eigenstate Thermalization Hypothesis (ETH) ceases to hold. This indicates a transition beyond which thermalization no longer accurately describes the system's behavior. Bottom panel: same as the top panel, but   for the entanglement entropy  between   the central site and the rest of the chain. Insets provide  zoomed-views of the central energy region, highlighting the  reduction  of the  ranges  with increasing system size.}
    \label{fig:MAxMin}
\end{figure}

To further support the validity of  the Eigenstate Thermalization Hypothesis\cite{Rigol(2010)}, Figures \ref{fig:StandarDeviation} and \ref{fig:MAxMin} present two complementary measures of eigenstate-to-eigenstate fluctuations.  Specifically, Fig.  \ref{fig:StandarDeviation} shows the standard deviation, while Fig. \ref{fig:MAxMin} displays the range (maximum minus minimum) of the entanglement entropy and the number of particles at the central site. Both quantities are computed   using  moving windows of 200 consecutive  eigenstates and are plotted as   functions of the mean energy within each window. These analysis are performed for four system sizes:  $N=M=7,8,9$ and $10$. For both observables and both fluctuation measures, a clear trend is observed: in the low-energy regime ($E/N\lesssim 1.1$), the magnitude of fluctuations decreases as the system size increases. This behavior provides strong evidence in support of the ETH within the same energy interval where  signatures of quantum chaos were  previously identified in the model  (see Fig.\ref{fig:r_promedio}).

\subsection{Dynamical thermalization of Fock states}
\label{sec:dynamical fock}
Having established the energy interval in which the Eigenstate Thermalization Hypothesis (ETH) holds, this subsection examines whether initial states defined in the occupation basis evolve toward thermal equilibrium. To this end, the long-time (equilibrium) values of the observables are compared with their corresponding microcanonical averages, in order to assess whether these initial states exhibit thermalization behavior within the IAA model.

Consider an isolated quantum system initially prepared in a nonstationary state with a well-defined mean energy. In such a scenario,  an observable is said to  thermalize if, during  the system's time evolution, its expectation value  gradually converges toward  the microcanonical prediction and  remains close to it for most later  times. Notably, the distinction between pure and mixed initial states plays no essential role in determining whether thermalization occurs \cite{Alessio(2016)}. 
To describe the time evolution of such a system, the initial state can be expanded in terms of the eigenbasis $\{\ket{\phi_l}\}$ of the Hamiltonian $\hat{H}$, as $\ket{\Psi(0)} = \sum_l c_l \ket{\phi_l}$.

At time t, the state evolves as $\ket{\Psi(t)} = e^{-i\hat{H}t}\ket{\Psi(0)} = \sum_l c_l e^{-iE_lt}\ket{\phi_l}$, where $E_l$ denotes the energy corresponding to each eigenstate. The time evolution of any observable $\hat{A}$ is described by the expression:
\begin{equation}
\begin{split}
\hat{A}(t) &= \braket{\Psi(t)|\hat{A}|\Psi(t)}
= \sum_m |c_m|^2 A_{m,m}\\
& \qquad +\sum_{m,n\neq m}c_m^* c_n e^{i(E_m-E_n)t}A_{m,n},
\label{eq:observable}
\end{split}
\end{equation}
where $A_{m,n} = \braket{\phi_m|\hat{A}|\phi_n}$. In the long-time average, the second sum in Eq.~(\ref{eq:observable}) typically vanishes -assuming the  absence of  degeneracies or that the number of any existing degeneracies  is nonextensive.   Consequently, we are left with the sum of the diagonal elements of $\hat{A}$ weighted by $|c_m^2|$. 

Thermalization of the observable $\hat{A}$ implies that the first term in Eq.~(\ref{eq:observable}) converges to the microcanonical average. This microcanonical average  is defined as the average value of the observable over all  eigenstates whose energies fall  within a specified window around an energy  $E$.  Formally, it is given by
\begin{equation}
\langle \hat{A} \rangle_{\text{micro}} = \frac{1}{\mathcal{N}(E)} \sum_{E_i \in [E \pm \Delta E]} 
A_{i,i}
%\langle \phi_i | \hat{A} | \phi_i \rangle
\label{eq:microav}
\end{equation}
where $\mathcal{N}(E)$ denotes  the number of eigenstates $\ket{\phi_i}$ with energies in the interval $[E-\Delta E, E + \Delta E]$, and 
$A_{i,i}$
%$\braket{\phi_i|\hat{A}|\phi_i}$ 
represents  the expectation value of the observables $\hat{A}$ in the corresponding Hamiltonian eigenstate, as defined above.  

In simpler terms, the microcanonical average corresponds to the mean value of a quantum observable, evaluated over all Hamiltonian eigenstates whose eigenenergies fall within a narrow interval centered around a reference energy E. This statistical average serves to characterize the behavior of the observable at equilibrium, under the assumption that the system is isolated and possesses a fixed total energy.
In our analysis, the microcanonical average was computed using a spectral window centered at energy $E$, including the 30 nearest eigenstates above and below this reference energy. The window was systematically shifted across the spectrum in single-state increments to capture equilibrium behavior throughout the system.
In what follows, equilibrium values of observables were obtained  for $t = 10^4$, and these results have been averaged over $40$ independent disorder realizations of the model.

\begin{figure}[h]
\centering
\begin{subfigure}
\centering
\includegraphics[scale=0.5]{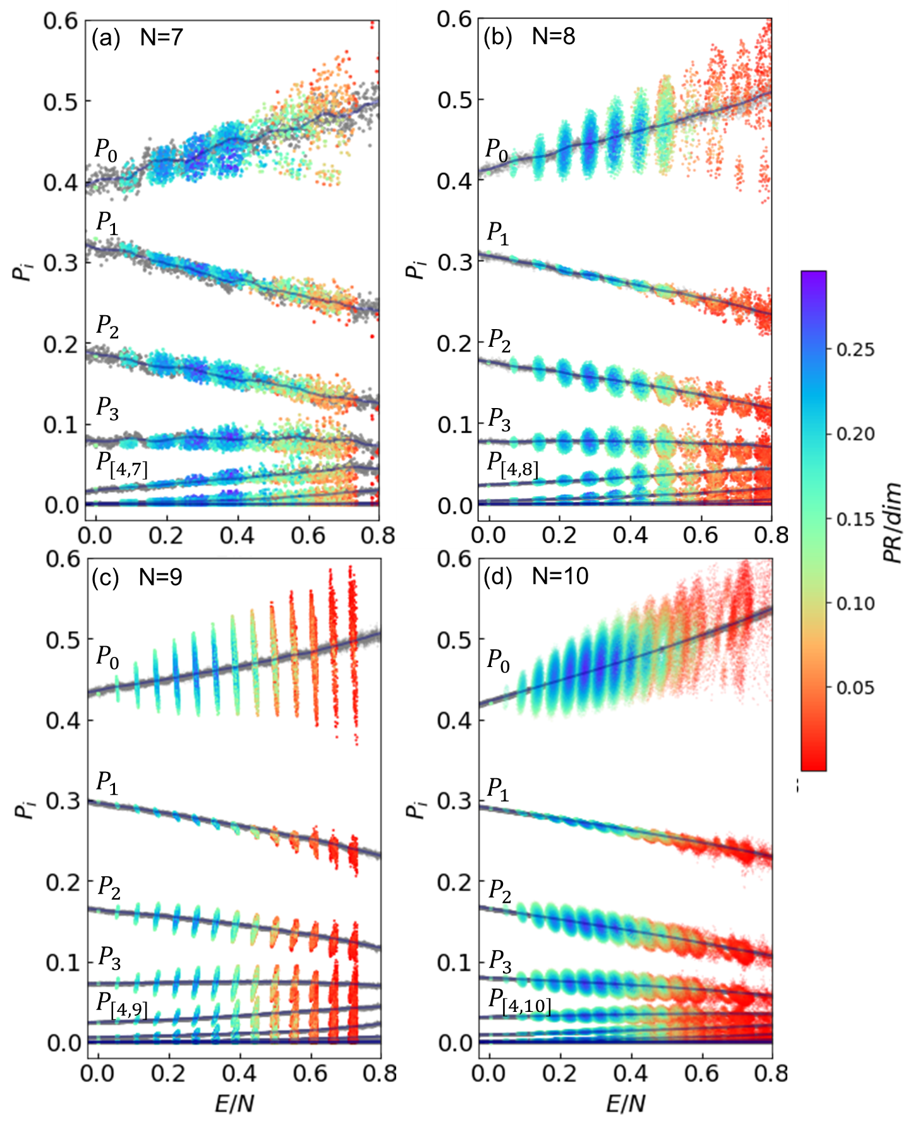}
\end{subfigure}
\caption{ Colored dots:  Equilibrium values (at $t\approx 10^4$) of the occupation probabilities at  the central site,  for  initial Fock states, plotted against their mean energy per particle ($E/N$). The  color scale indicates  the ratio  between the  participation ratio (PR) of the occupation states in the energy eigenbasis and the dimension of the Hilbert space.   Gray dots:  Expectation values of the same  occupation probabilities in the energy eigenstates. Dark blue line represents the  microcanonical average, computed as the mean value over  rolling windows of 30 energy eigenstates. Results are shown for four  system sizes:  $N=M=7,8,9$ and $10$.}\label{fig:pis_ex_eth_zoomout}
\end{figure}

Fig. \ref{fig:pis_ex_eth_zoomout} presents a comparison between the equilibrium  occupation probabilities $P_i$ at the central site  of the chain  and the corresponding  values obtained from individual  eigenstates (shown as gray dots). The microcanonical average is represented  by a continuous dark blue line, while  the Participation Ratio (PR)  corresponding to each  occupation state is encoded using a color scale. 

The equilibrium values clearly  organize into several clusters, which contain a number of states that increases  with  energy. These  clusters correspond to distinct values  of the crowding parameter. The leftmost cluster,  which consists of  a single point associated  with  $C=1$, represents  the Mott state  -characterized by one atom per site.  

Importantly, clusters at lower energies exhibit occupation probabilities that closely match the microcanonical average. As energy increases, the clusters display broader dispersions, although the rate of dispersion varies depending on the particular $P_i$ under consideration. For instance, $P_1$ exhibits significantly lower dispersion than other occupation probabilities.
Notably, the clusters associated with $P_1$ align almost perfectly with the eigenstate probabilities within the energy range $E/N \in [0, 0.5]$, across all four system sizes studied. Overall, the dispersion of values grows with the crowding parameter (and  consequently energy), and within each cluster, states characterized by lower PR values tend to show greater deviations from the microcanonical prediction.

\begin{figure}
\centering
\begin{subfigure}
\centering
\centering
\includegraphics[scale=0.45]{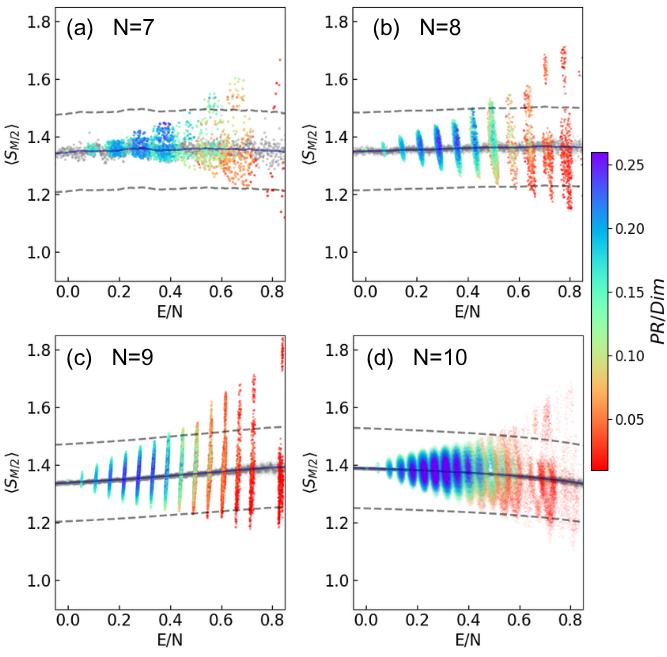}
\end{subfigure}
\caption{ Same as Fig.\ref{fig:pis_ex_eth_zoomout}, but  for the entanglement entropy of the central site  with respect to all  other sites in the chain. Gray dashed lines indicate a $10 \%$  deviation  from the  microcanonical average.}
\label{fig:pis_ex_eth}
\end{figure}

\begin{figure}[h!]
\centering
\begin{subfigure}
\centering
\subfigure{\includegraphics[width=0.45\textwidth ]{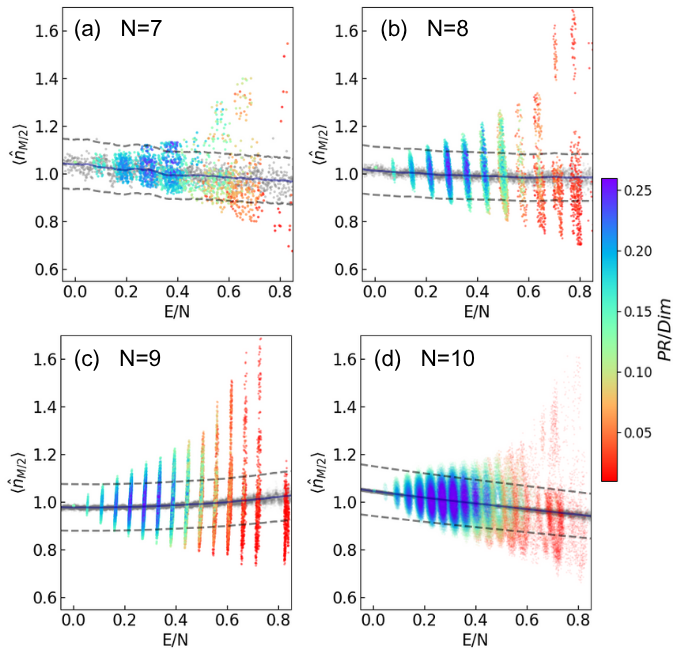}}\end{subfigure}
\caption{
Same as Fig.\ref{fig:pis_ex_eth_zoomout}, but  for the  particle  number at  the central  site of the chain. Gray dashed lines indicate a $10 \%$  deviation  from the  microcanonical average.
} \label{fig:s1n1_fock_zoomout}
\end{figure}

We now examine  the equilibrium values of the two observables previously analyzed for the energy eigenstates: the entanglement entropy  (Fig.\ref{fig:pis_ex_eth})  and  the number of particles  at the central site of the chain (Fig.\ref{fig:s1n1_fock_zoomout}).
To gain a comprehensive understanding of the relaxation  behavior of   $S_{M/2}^{(q)}$  and $n_{M/2}^{(q)}$,  we plot their  average values -computed over  40 Hamiltonian realizations- for all initial Fock states with energies in  the chaotic regime $E/N\in [0,0.8]$. 
Figures \ref{fig:pis_ex_eth} and  \ref{fig:s1n1_fock_zoomout} display these relaxation values for four system sizes ($N=M=7,8,9$ and $10$). 
The  values are color-coded using  a rainbow scale to represent  the PR of each  initial states. For reference,  ensemble  averages  of the corresponding  observables calculated over  energy  eigenstates are shown as  gray dots in the background. The microcanonical average, defined in Eq. (\ref{eq:microav}), is plotted as a solid line, while dashed lines indicate $\pm 10 \%$ deviations.

Similar to the occupation probabilities, the relaxation values of the observables form distinct clusters, each corresponding to  a specific value  of the  crowding parameter. Starting with   $C=1$ on the left,    the crowding parameter  of the associated Fock states increases from left to right.  In all panels, the relaxation value of the Mott state ($C=1$)  falls within the band of values coming from the energy eigenstates, indicating that   the ETH holds  for this state. This initial state was the subject of extensive experimental investigation in Ref.~\cite{Lukin(2019)}. 

For  states with higher energies,  the ETH is not perfectly satisfied across all  states; however the relaxation values within each cluster  follow    the trend established  by the  ensemble averages of the energy eigenstates. Across  all system sizes, the majority  of  equilibration values  for  the entanglement entropy fall within the $10 \%$ tolerance  margin of the microcanonical average for energies $E/N < 0.5$. In contrast, equilibrium values for  the boson number  show greater deviations, with  this tolerance range narrowing   to $E/N< 0.2$.      
The dispersion of the relaxation values within  each cluster increases with  energy, but, in general,  the equilibration values of Fock states with high PR present  lower dispersion and tend to concentrate near the cluster centers.  Conversely, equilibration values of  states with low PR occupy a broader region  within the respective cluster and display markedly  larger variability.

\subsection{Quantifying the lack of  thermalization.}
\label{sec:lack therm}
From the results presented in  the previous subsection, it is evident that  relaxation values of Fock basis states with low PR  tend to deviate more strongly from thermal averages  compared to those  with high PR. In this subsection, we  assess whether   these deviations are statistically  significant using   the trace distance --a metric  that is independent of any specific observable.

\begin{figure}[h!]
    \centering
    \includegraphics[scale=0.5]{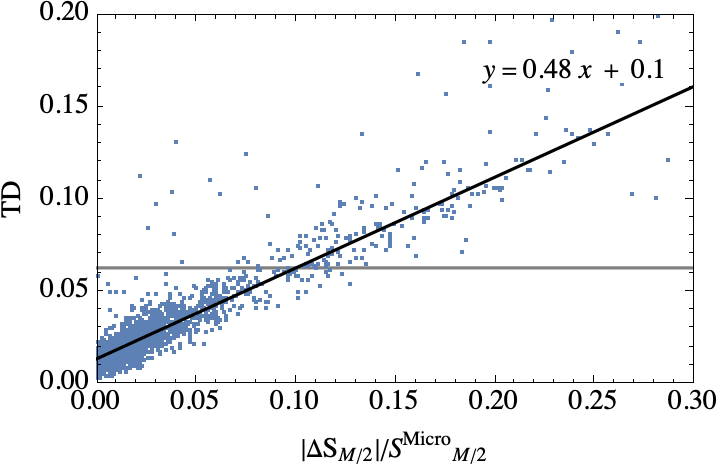}
    \includegraphics[scale=0.5]{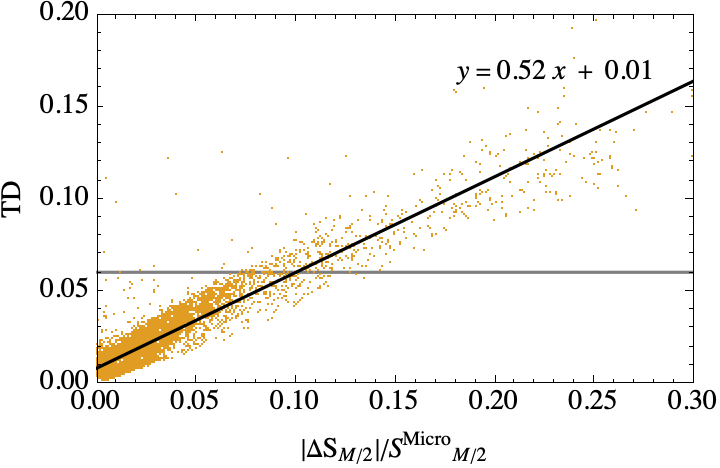}
    \includegraphics[scale=0.5]{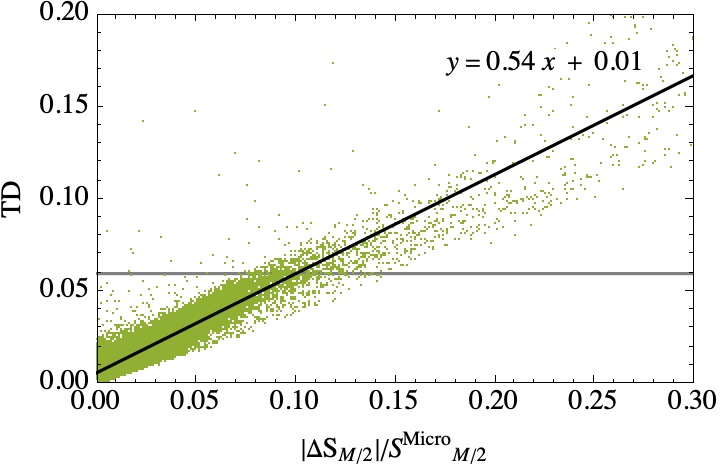}
    \caption{ Trace distance between the microcanonical reduced  density matrix and the reduced density matrix of states evolved unitarily (at $t=10^4$) from initial Fock states, Eq.(\ref{eq:TDsimple}),  plotted against the relative entropy deviation $\Delta S_{M/2}/ S_{M/2}^\text{Micro}$, for three different system sizes: top panel is for  $N=M=7$, middle panel  for  $N=M=8$ and bottom panel  for $N=M=9$. A linear  correlation (continuous black line) is observed between these two quantities. The gray line marks  the trace distance  value $TD\approx 0.07$, corresponding to a  10\%  deviation  ($\Delta S_{M/2} / S_{M/2}^{\text{micro}} = 0.1$) between the microcanonical entropy and the equilibration entropy of the initial Fock states.}
    \label{fig:TDvsDELTA}
\end{figure}

\begin{figure}
    \centering
    \includegraphics[scale=0.4]{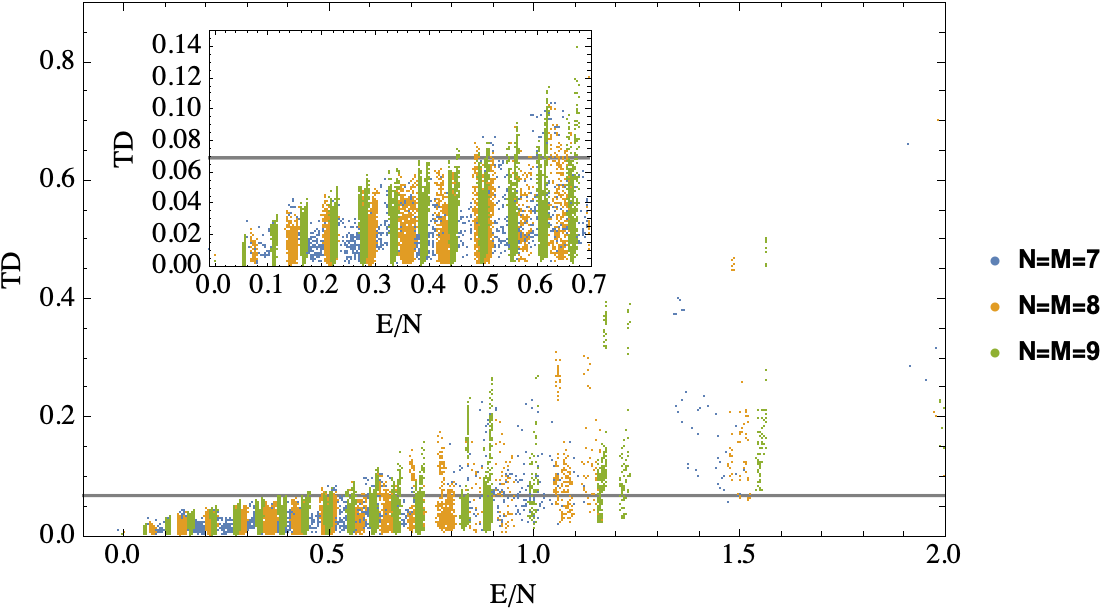}
    \caption{
 Trace  distance, as in Figure \ref{fig:TDvsDELTA}, 
 plotted as a function of the energy of the initial Fock  states for three different system sizes. The horizontal gray line marks TD$=0.07$. States with trace distance below this threshold reach  equilibrium values that closely match the thermal (microcanonical) values,  within  a 10 \%   tolerance. }
    \label{fig:TDvsE}
\end{figure}

\begin{figure*}
    \centering
    \includegraphics[scale=0.45]{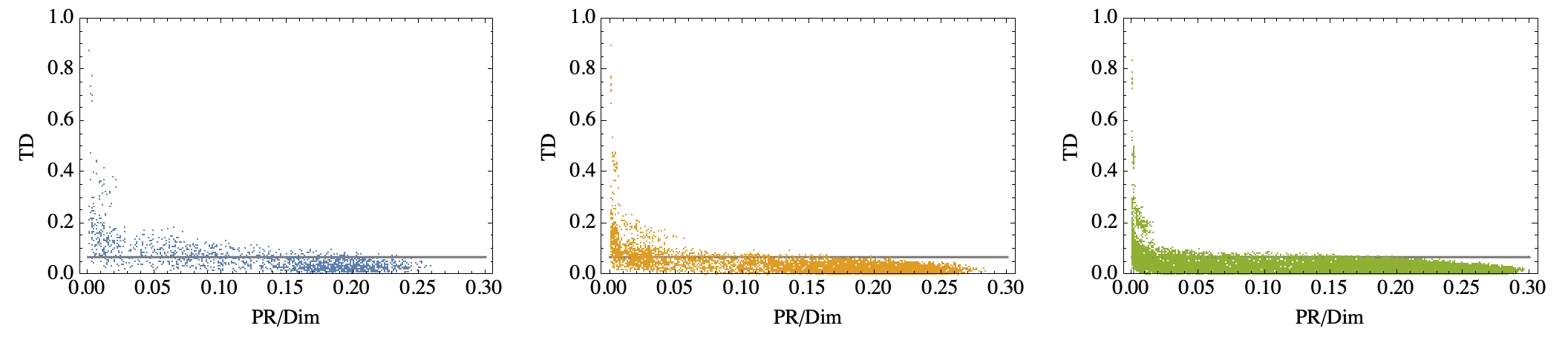}
    \caption{ 
   Trace  distance, as in Figure \ref{fig:TDvsDELTA}, plotted   against   the  normalized participation ratio (PR/Dim) of the Fock states with  respect to the energy eigenbasis. Results are shown for three different system sizes:   $N=M=7$ (left),  $N=M=8$ (middle) and  $N=M=9$ (right). States with low PR exhibit high trace distance  (TD) values, while those with high PR tend to show  lower TD values.  The horizontal gray line marks  the threshold TD$=0.07$, distinguishing whether the equilibrium state reached by initial Fock-basis states is  thermal (TD<0.07) or non-thermal  (TD>0.07).  }
    \label{fig:TDvsPR}
\end{figure*}

To quantify the thermal deviation  of the equilibrium state reached  by  initial  states in the Fock  basis, we consider two complementary measures: 1) the relative deviation of the entanglement entropy at the central site of the chain with  respect to its  thermal average, and 2) the trace distance between  the    microcanonical density matrix at the central site  $\rho_A^{mc}$ and  the reduced density matrix $\rho_A$ of  the central site  obtained  from  unitarily evolving initial Fock states. Formal definitions of  both quantities are presented below. 

The relative deviation of the entanglement entropy from its  thermal value is defined as  
\begin{equation}
\frac{|\Delta S_{M/2}|}{S_{M/2 \text{micro}}}=\frac{|S_{M/2}-S_{M/2 \text{micro}}|}{S_{M/2 \text{micro}}},
\end{equation}
where $S_{M/2 \text{micro}}$ denotes the entropy of the central site  
calculated within  the microcanonical ensemble [see  Eq.~(\ref{eq:microav})], and $S_{M/2}$ represents the equilibrium   entropy of the initial Fock state. 

Additionally,  we evaluate the Trace Distance (TD) between  the two density matrices at the central site given by 
\begin{equation}
\text{TD(A)} = \frac{1}{2} \text{Tr} \left| \rho_{A}^{\text{mc}} - \rho_A \right|.
\end{equation}
Since our observables are restricted to  a  single site, the trace distance simplifies to 
\begin{equation}
\text{TD(A)} = \frac{1}{2} \sum_i^M \left| P_{i, \text{micro}} - P_i^{(A)} \right|.
\label{eq:TDsimple}
\end{equation}
When TD is close to zero, the two matrices are similar, and the system is said to thermalize.  In contrast, values of  TD approaching  one indicate non-thermal behavior. In the following, we establish  a reference threshold for  TD, enabling  us to discriminate  between thermal and non-thermal states. 

Fig.\ref{fig:TDvsDELTA} displays   the trace distance plotted against   the relative deviation of the entanglement entropy  for all initial states in the Fock basis,  for system sizes  $N=M=7,8$ and $9$. As  expected,  the figure reveals  a simple correlation between the two quantities.  For all three system sizes considered, a linear relationship is observed,  TD~$\propto |\Delta S_1|/S_{1 \text{micro}}$. From these results, we infer  that thermal deviations in  equilibration values  below $10\%$ correspond to TD$<0.07$. Consequently, this  value can be used as a threshold to distinguish  thermalizing states (TD$<0.07$) from non-thermalizing ones (TD$>0.07$).  It is important  to emphasize that  the trace distance  depends solely on the quantum  states involved and is independent of any specific observable chosen. Therefore, a small trace distance ensures thermalization across all local observables.

Figs. \ref{fig:TDvsE} and \ref{fig:TDvsPR} illustrate the behavior of the equilibrium TD for all initial states in the Fock  basis. In Fig. \ref{fig:TDvsE}, we present the TD values as a function of energy for three different system sizes.  For states  with  $E/N < 0.5$,  the majority  of  TD values  fall below  0.07, indicating that  these states predominnatly  thermalize. Conversely,  at  higher energy, an increasing fraction of Fock states fail to thermalize. 

Notably, across all system size,  the value $E/N \approx 0.5$ marks the onset  of pronounced  proliferation of non-thermalizing states. This transition closely  aligns with the shift from chaotic to regular spectral behavior,  previously identified in  Fig.\ref{fig:r_promedio}. The  breakdown of  thermalization at high energies is attributed to the loss of chaotic dynamics and the emergence of boson localization, where particles become confined within individual potential wells due to disorder. 

Fig. \ref{fig:TDvsPR}  presents again   the TD but now as a function of the normalized participation ratio, PR/Dim,  for  the states in the Fock basis, across three system sizes.  The results indicate that  states with high PR/Dim>$0.20$ (i.e., exceeding one-fifth of the Hilbert space dimension) consistently exhibit trace distances below $0.07$, suggesting that   they thermalize.  As  the value of PR/Dim decreases,  the  fraction of  states with  TD$>0.07$,  increases, signaling a growing prevalence of non-thermalizing behavior. Based on this criterion, one can conclude that a participation ratio larger  than Dim/5 is a sufficient condition for thermalization. However,  states with lower  PR values do not uniformly fail to thermalize, a portion of them still exhibit thermal behavior despite falling below this threshold.

\section{\label{sec:conclusion}Conclusions}
In this work, we performed  a detailed  numerical analysis of the interacting Aubry-André model to assess the thermalization behavior in  systems initialized  from   individual Fock  states. Spectral analysis revealed  a broad  energy regime characterized by quantum chaos, within which the  system's eigenstates conform  to the Eigenstate Thermalization Hypothesis (ETH). This was evidenced by smooth energy dependence of  local observable expectation values  and diminishing eigenstates fluctuations   with increasing  system size. 

Our central finding is that thermalization of Fock states is critically governed by their structure in the energy eigenbasis. Specifically, we demonstrated a strong correlation between a state's delocalization --quantified by the Participation Ratio (PR)-- and its approach to thermal equilibrium. States with high PR are sufficiently extended in the eigenbasis to explore the full energy shell, evolving toward microcanonical predictions. In contrast, low-PR states remain localized, retain memory of initial conditions, and exhibit significant deviations from thermal averages. Trace distance analysis confirmed this relationship, establishing high PR as a sufficient condition for thermalization.
These results underscore a direct link between eigenbasis delocalization and dynamical equilibration, offering a predictive framework for identifying thermalizing states in disordered quantum systems.

The observed correlation between low PR and non-thermal behavior can arise from multiple mechanisms that may act simultaneously. At high energies ($1.0\leq E/N$), many-body localization and the accompanying regular dynamics inhibit thermalization of Fock states. Even within the chaotic regime ($0.5\leq E/N<1.0$), proximity to the mobility edge may introduce localized features and consequent non-thermal behavior.  For Fock states deeply embedded in the chaotic domain ($0\leq E/N<0.5$), ETH is generally satisfied within a certain  tolerance, yet a  low PR still correlates with enhanced fluctuations in equilibration values. Possible contributing factors  include border effects, quantum scars, and sparse initial boson  configurations \cite{Pausch_2025_seed_erg} that hinder ergodic exploration. Further investigations (work in progress) employing systems with periodic boundary conditions, along with  refined metrics for quantifying the sparsity of initial Fock states, are necessary to elucidate the underlying mechanisms responsible for the observed correlation between participation ratio and thermalization.
           
\acknowledgments 
We acknowledge the support of the Computation Center - ICN to develop many of the results presented in this work, in particular to Enrique Palacios, Luciano Díaz, and Eduardo Murrieta. 
This work received partial financial support form DGAPA- UNAM project IN109523.

%\clearpage
\bibliographystyle{unsrt}
\bibliography{bibliography}
\end{document}